\documentclass[traditabstract]{aa}
\usepackage{txfonts}
\usepackage{graphicx}

\def\deg{^{\circ}}
\newcommand{\ve}[1]{{\rm\bf {#1}}}

\begin{document}

\title{Inferring the magnetic field vector in the quiet Sun}
\subtitle{I. Photon noise and Selection Criteria}

\author{J.M.~Borrero\inst{1} \and P. Kobel\inst{2}}
\institute{Kiepenheuer-Institut f\"ur Sonnenphysik, Sch\"oneckstr. 6, D-79110, Freiburg, Germany. \email{borrero@kis.uni-freiburg.de}
 \and Max-Planck Institut f\"ur Sonnensystemforschung, Max-Planck Str. 2, Katlenburg-Lindau, 37191, Germany. \email{kobel@mps.mpg.de}}
\date{Recieved / Accepted}

\begin{abstract}
{In the past, spectropolarimetric data from Hinode/SP has been employed to infer the distribution of the magnetic
field vector in the quiet Sun. While some authors have found predominantly horizontal magnetic
fields, others favor an isotropic distribution. In this paper, we investigate whether it is actually 
possible to accurately retrieve the magnetic field vector in regions with very low polarization signals (e.g:
internetwork), employing the \ion{Fe}{I} line pair at 6300 {\AA}. We first perform inversions 
of the Stokes vector observed with Hinode/SP in the quiet Sun at disk center in order to confirm the distributions
retrieved by other authors. We then carry out several Monte-Carlo simulations with synthetic data 
where we show that the observed distribution of the magnetic field vector can be explained in terms of purely 
vertical ($\gamma=0\deg$) and weak fields ($\bar{B}<20$ G), that are misinterpreted by the analysis technique
(Stokes inversion code) as being horizontal ($\gamma \approx 90\deg$) and stronger ($\bar{B} \approx 100$ G), due 
to the effect of the photon noise. This casts doubts as to whether previous results, presenting the distributions
 for the magnetic field vector peaking at $\gamma=90\deg$ and $\bar{B}=100$ G, are actually correct. We propose that 
an accurate determination of the magnetic field vector can be achieved by decreasing the photon noise to a point where 
most of the observed profiles posses Stokes $Q$ or $U$ profiles that are above the noise level. Unfortunately, for noise 
levels as low as $2.8\times 10^{-4}$ only 30 \% of the observed region with Hinode/SP have strong enough $Q$ or $U$ 
signals, implying that the magnetic field vector remains unknown in the rest of the internetwork.}
\end{abstract}

\keywords{Magnetic fields -- Line: profiles -- Polarization -- Sun: photosphere -- Sun: surface magnetism}
\maketitle

\section{Introduction}

The magnetic field in the quiet Sun internetwork has been subject to an intense debate over the years. Its study has been
motivated by the fact that much of solar magnetic flux could be contained in this region (S\'anchez Almeida 2004), which 
has important consequences in our understanding of the solar activity cycle (Spruit 2003; Solanki et al. 2006). In 
addition, quiet Sun magnetic fields are subject to the strong influence of
the granular convection, giving raise to a rich variety of phenomena such as magnetic flux emergence (Centeno et al. 2007;
Mart{\'\i}nez Gonz\'alez \& Bellot Rubio 2009; Danilovic et al. 2010), convective collapse (Bellot Rubio et al. 2001; Nagata et 
al. 2008; Fischer et al. 2009), supersonic velocities (Socas-Navarro \& Manso-Sainz 2005; Shimizu et al. 2007; Bellot Rubio 2009; 
Borrero et al. 2010) and so forth, that play an important role on the heating of the upper layers of the solar atmosphere (Schrijver et al. 
1997, 1998; Schrijver \& Title 2003).

These studies have triggered the development of instruments with ever-increasing spatial resolution, observations in different 
spectral windows (Khomenko et al. 2003; Dom{\'\i}nguez Cerde\~na et al. 2006; Socas-Navarro et al. 2008), and using spectral lines subject 
to the Zeeman and Hanle effect (Manso Sainz et al. 2004, Trujillo Bueno et al. 2004; S\'anchez Almeida, J. 2005). A mayor step 
has been accomplished with the deployment of the Japanese spacecraft Hinode. The spectropolarimeter (SP) attached to the 
Solar Optical Telescope (SOT; Tsuneta et al. 2008) on-board Hinode allows to achieve continuous observations of the solar 
atmosphere with an extremely high-spatial resolution. This advantage has favored the use of data from this 
instrument (Ichimoto et al. 2008) in the investigations of the quiet Sun magnetism over the last few years. The analysis of this
data has revealed that a large portion of the magnetic flux in the internetwork presents itself in the form of horizontal magnetic fields
(Orozco Su\'arez et al. 2007a, Lites et al. 2008). Other works, also analyzing Hinode/SP data favor an isotropic (e.g: turbulent)
distribution of the magnetic field vector (Asensio Ramos 2009; Stenflo 2010). Motivated by this controversy we have revisited
these results with the hope of offering possible solutions. In Section 2 we describe our observations and data-sets from the
Hinode/SP instrument. Section 3 describes the analysis method and how we infer the magnetic field vector from spectropolarimetric
observations. Section 4 presents our results and compares them with previous works, whereas Sections 5 and 6 are fully devoted 
to study the effects that photon noise and the different selection criteria applied to the data, have in the inferred distribution
 of the magnetic field. Finally, Section 7 summarizes our findings.

\begin{figure*}
\begin{center}
\includegraphics[width=16cm]{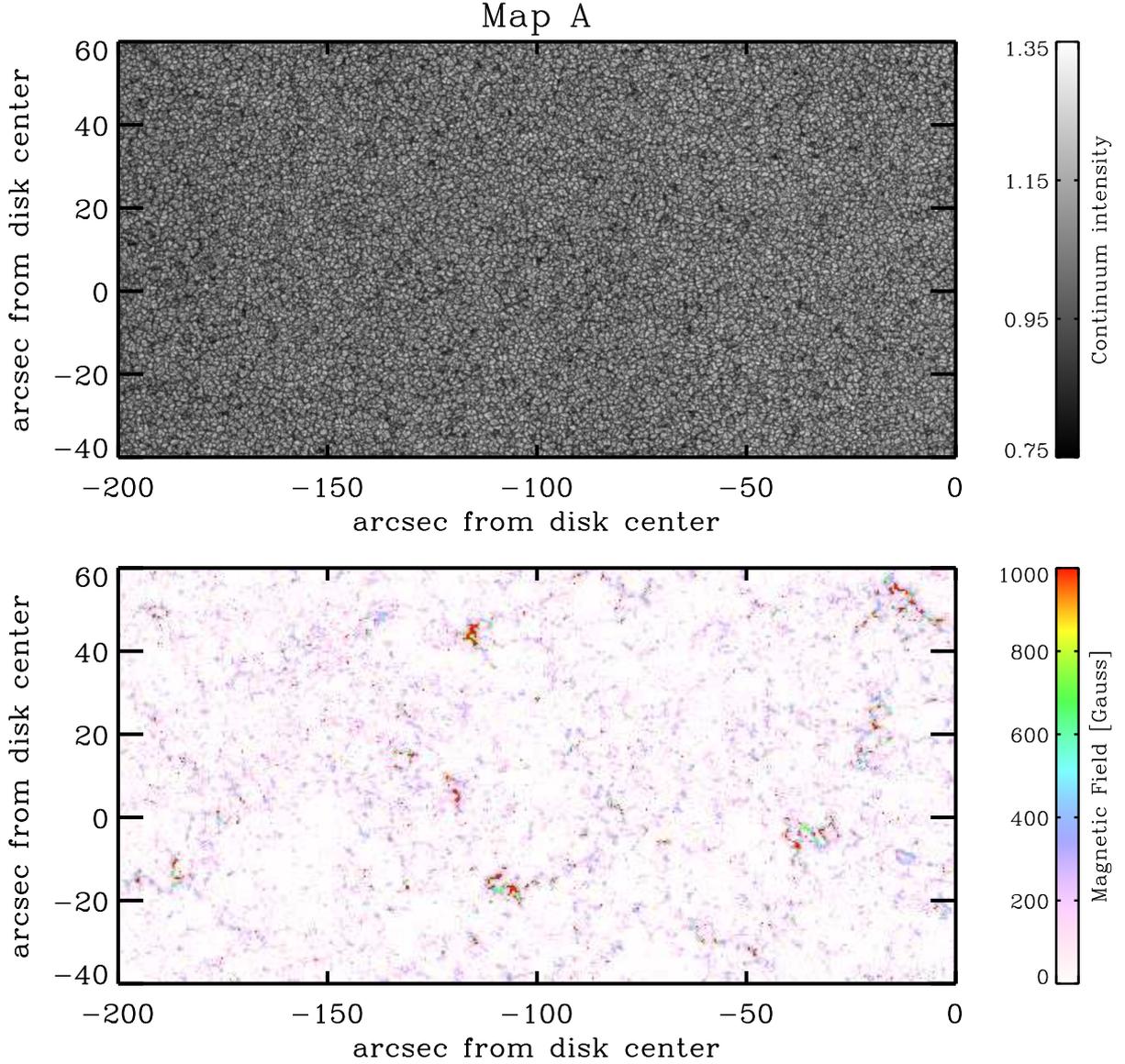}
\caption{Observed continuum intensity (upper panel) and mean magnetic field strength at each pixel
inferred from the inversion of the Stokes profiles (lower panel). This region is only a subsection of the full
field-of-view. It was chosen to match approximately Fig.~2 in OS07a. See Section 2.1 for details.}
\end{center}
\end{figure*}

\section{Observations}

The data used in this work correspond to spectropolarimetric observations (full Stokes vector: $I$, $Q$, $U$, $V$)
around the \ion{Fe}{I} line pair at 6301.5 {\AA} and 6302.5 {\AA}, recorded with the spectropolarimeter
on-board the Japanese satellite Hinode (Lites et al. 2001; Kosugi et al. 2007; Tsuneta et al. 2008;
Suematsu et al. 2008; Ichimoto et al. 2008). These two lines have Land\'e factors
of $g_{\rm eff} =1.67$ and $g=2.5$. The image stabilization system in Hinode (Shimizu et al. 2008)
allows the slit spectropolarimeter to achieve a spatial resolution of 0.32". Here we analyze several data sets 
corresponding to quiet Sun observations. The maps employed are the sames as the ones already analyzed by Orozco Su\'arez 
et al. (2007a; 2007b) and Lites et al. (2008). Hereafter these three papers will be referred to as: OS07a, OS07b 
and LIT08. Further analysis of some subsections of the same dataset have also been carried out by Asensio Ramos (2009), 
Mart{\'\i}nez Gonz\'alez et al. (2010) and Stenflo (2010). In the following subsections we briefly describe each selected
map. A summary of the main their properties can be also found in Table 1.

\subsection{Normal Hinode/SP map}

This map contains 2047$\times$1024 pixels with a pixel size of 0.15" and 0.16" in the horizontal and
vertical directions, respectively.  The field of view is centered at (-53.1",7.9") from disk center
($\mu=0.998$). The exposure time at each slit position is 4.8 s, yielding a noise level of $\sigma_{\rm s}\approx 10^{-3}$
(in units of the average quiet Sun intensity). Here, the index $s$ refers to any of the four components
of the Stokes vector. Hereafter we will refer to this map as \emph{Map A}. The
data was recorded in March 10, 2007, from 11:37 UT to 14:36 UT. A map of the continuum intensity and magnetic 
field strength (retrieved by the inversion; Sect.~4) in a subfield of this map is displayed in Figure 1. The subfield has been chosen to 
coincide with Figure 2 in OS07a.

\subsection{Deep Hinode/SP map}

This map contains 727$\times$1024 pixels with a pixel size of 0.15" and 0.16" in the horizontal and
vertical directions, respectively. In this case, the slit did not scan across the solar surface, but instead it
was kept at the same location (-22.4",7.8") on the sun from 00:20 UT to 02:19 UT on February 27, 2007.
The exposure time at each slit position is 9.6 s, yielding a noise level of $\sigma_{\rm s} \approx 7.5\times 10^{-4}$ 
(in units of the average quiet Sun intensity).  Hereafter we will refer to this map as \emph{Map B}. Figure 2 
shows a map of the continuum intensity and magnetic field strength.

\subsection{Low noise Hinode/SP map}

In order to further decrease the noise level we have averaged every 7 slit exposures in Map B
producing a new map (referred to as \emph{Map C}) that contains 103$\times$1024 pixels, with an effective
exposure time of 67.2 s. The increased integration time yields a noise of level $\sigma_{\rm s} \approx 2.75\times 10^{-4}$ in
units of the continuum intensity. 

\begin{table}
\caption{Summary of employed datasets.}
\begin{tabular}{cccccccc}
\hline
Map & Date-Obs & Time-Obs & $\sigma_{quv}$ [10$^{-3}$/I$_c$] & $\chi_{\alpha}^2$ & $\chi_{\beta}^2$ \\
\hline
A & Mar 10, 2007 & 11:37$-$14:36 & 1.04 & 0.95 & 0.88 \\
B & Feb 27, 2007 & 00:20$-$02:19 & 0.75 & 1.48 & 1.34 \\
C & Feb 27, 2007 & 00:20$-$02:19 & 0.28 & 0.98 & 0.90 \\
\end{tabular}
\end{table}

\begin{figure*}
\begin{center}
\begin{tabular}{cc}
\includegraphics[width=9cm]{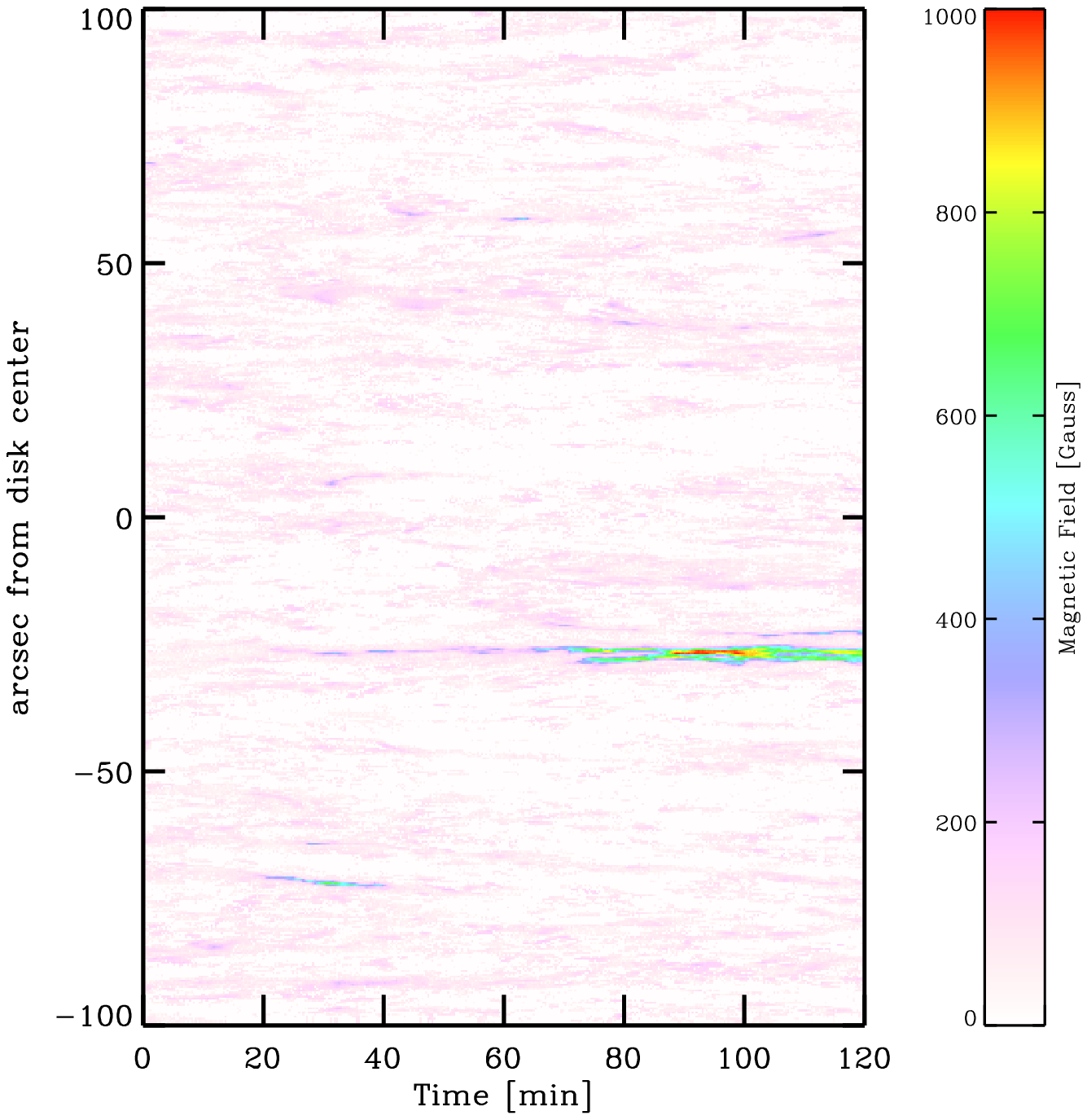} &
\includegraphics[width=9cm]{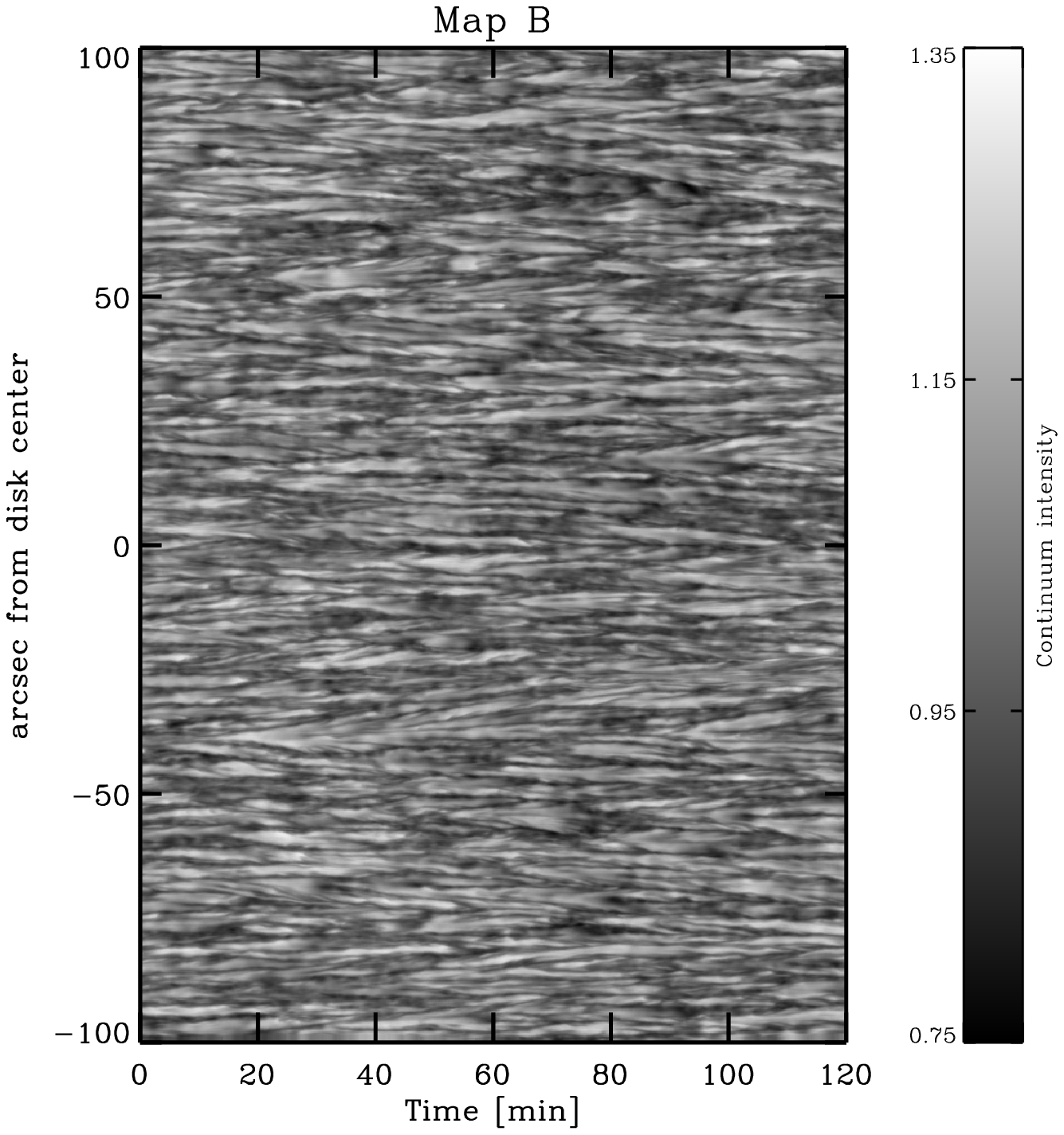}
\end{tabular}
\caption{Observed continuum intensity (upper panel) and mean magnetic field strength at each pixel
inferred from the inversion of the Stokes profiles (lower panel). Note that the spectrograph's slit was located
always at the same position on the Sun, therefore the horizontal axis refers to time. See Section 2.2 for details.}
\end{center}
\end{figure*}

\section{Inversion of Stokes profiles and magnetic field vector retrieval}

The magnetic field vector was retrieved from the Stokes vector using the VFISV (Very Fast Inversion
of the Stokes Vector; Borrero et al. 2010) inversion code\footnote{This inversion code is freely
available for download at the Community Spectro-Polarimetric Analysis Center (CSAC; Lites et al. 2007), 
which is an initiative at the High Altitude Observatory and National Center for Atmospheric Research (NCAR):
http://www.csac.hao.ucar.edu/}. We employ a parametrized atmospheric model that 
contains a magnetic and a non-magnetic atmosphere next to each other. The magnetic atmosphere is
characterized by the following parameters: magnetic field strength $B$, inclination of the magnetic field with 
respect to the observer $\gamma$, azimuthal angle of the magnetic field in the plane perpendicular to the 
observer $\phi$, line-of-sight velocity $V_{\rm LOS}$, damping parameter $a$, ratio of the absorption coefficient 
between the line-core and the line-continuum $\eta_0$, Doppler-width $\Delta \lambda_{\rm Dop}$, source function
at the observer's location $S_0$, and the gradient of the source function with optical depth $S_1$. The non-magnetic
atmosphere has one single parameter: fractional area of the pixel filled with non-magnetic plasma  
(aka non-magnetic \emph{nm} filling factor) $\alpha_{\rm nm}$. Using initial values for these parameters,
VFISV obtains the theoretical Stokes vector $\ve{I}_{\rm tot}=(I,Q,U,V)$ as:

\begin{equation}
\ve{I}_{\rm tot} = (1-\alpha_{\rm nm}) \ve{I}_{\rm mag} + \alpha_{\rm nm} \ve{I}_{\rm nm} \;,
\end{equation}

\noindent where $\ve{I}_{\rm mag}$ is the polarized contribution of the magnetized atmosphere, which is obtained
by solving the radiative transfer equation under the assumptions of the Milne-Eddington (M-E) approximation. 
$\ve{I}_{\rm nm}$ corresponds to the non-magnetic contribution, which is thus unpolarized ($Q_{\rm nm}=U_{\rm nm}=V_{\rm nm}=0$: 
see Section 3.1). The total theoretical Stokes vector $\ve{I}_{\rm tot}$ is then compared with the observed one using 
a merit function $\chi^2$, and by means of a non-linear least-squares-fitting algorithm (Levenberg-Marquardt method; 
Press et al. 1986) we iterate our initial 10 free parameters until a minimum in $\chi^2$ is reached. As an example, 
Figures 1 and 2 show the resulting magnetic field strength (from the inversion) for Maps A and B.

\subsection{Treatatment of the non-magnetic atmosphere}

In order to compare with OS07b, we have adopted for our inversions a treatment of the non-magnetic atmosphere 
that similar to theirs: a different $I_{\rm nm}$ profile is used in the inversion of each pixel in the map\footnote{Note 
that we have dropped the vector character of the stray-light profile (Equation 1) because of its non-polarized character. 
This means that $Q_{\rm nm}=U_{\rm nm}=V_{\rm nm}=0$ and therefore only the first component of the Stokes vector $I_{\rm nm}$
(total intensity) remains.}. $I_{\rm nm}$ is obtained for each pixel by averaging only the intensity 
profiles $I$ (polarization not included) from the neighboring pixels in a 1" square around the inverted one.
This method attempts to mimic a narrow spectrograph's point spread function. OS07b interprets the need for a non-magnetic
component as the reduction in the amplitude of the polarization signals (reduction $\propto [1-\alpha_{\rm nm}]$)
cause by diffraction. Danilovic et al. (2008) ascribes it as being caused by a small defocus in the SP instrument.

Another feature of our inversions is that we do not consider the free parameter $\alpha_{\rm nm}$ (Eq.~1), but rather
we use a different variable $\beta_{\rm nm}$, where:

\begin{equation}
\beta_{\rm nm} = \tanh^{-1}(2\alpha_{\rm nm}-1)
\end{equation}

Since $\alpha_{\rm nm}$ varies from 0 (pixel fully filled with the magnetic atmosphere) to 1 (pixel
fully filled with non-magnetized plasma), $\beta{\rm nm}$ varies (through Eq.~2) between $(-\infty,+\infty)$. 
The reason for employing $\beta_{\rm nm}$ instead of $\alpha_{\rm nm}$ is to prevent the inversion 
from reaching a local minimum in the $\chi^2$-hypersurface where $\alpha_{\rm nm}$ is either 0 or 1, 
resulting in an unrealistic number of pixels harboring these two values. The use of $\beta_{\rm nm}$
is supported by the fact that $\chi^2$ improves on average by about 10 \% when using it instead of
$\alpha_{\rm nm}$ (see 5th and 6th columns in Table 1).

Due to our treatment of the non-magnetic atmosphere some caution should be exercised when speaking about 
field strengths in this work. In our modeling of the Stokes parameters (Eq.~1) we assume that the light
we observe at each pixel has a polarized contribution from the pixel itself, plus a non-polarized contribution 
from the surrounding pixels. From this point of view we cannot talk about intrinsic field strengths but rather
about mean field strength inside the resolution element.

An alternative interpretation of Eq.~1 assumes that inside the resolution element there are two kinds
of atmospheres: a magnetized one and an non-magnetized one, where the magnetized one possesses only \emph{one}
magnetic field. From this point of view it is possible to refer to $B$ as the intrinsic field strength. 
However, Hinode/SP's resolution is not sufficient to fully resolve all magnetic structures in the quiet Sun. 
Therefore our treatment oversimplifies the problem, as we are assuming that the magnetic field is constant inside 
the magnetized atmosphere. In this respect, $B$ should be interpreted as the mean field strength 
in the magnetized atmosphere. This implies that both interpretations of Eq.~1, despite being different from
a physical point of view, are indistinguishable from each other. Because of this, in the following we will refer 
to the magnetic field as $\bar{B}$, in order to indicate its averaged nature.

It is important to mention that the results presented in this paper, in particular those in Section 5,
do not depend on the scheme employed to characterize the stray-light. As a matter of fact, the same results
are obtained if we consder a global\footnote{The word global referts to a stray-light profile $I_{\rm nm}$
that is common for all inverted pixels, and that is obtained as an average of all pixels, in the FOV, where
the polarization signals are below the noise level} stray-light profile, or even if it is completely neglected.

\subsection{One-line vs two-lines inversion}

One limitation of the VFISV inversion code is that it only allows to analyze the data from one spectral line
at a time. Hinode/SP records the full Stokes vector for the \ion{Fe}{I} line pair at 6301.5 {\AA} and 6302.5 {\AA} 
across 112 spectral positions. From these two lines the obvious choice, due to its larger 
sensitivity to magnetic fields (larger Land\'e factor; see Sect.~2), is \ion{Fe}{I} 6302.5. Therefore VFISV only makes use of
50 spectral points around the second spectral line. This results in an obvious loss of information, that will
yield slightly larger errors in the determination of our model parameters (Sect.~3; see also Orozco Su\'arez et al. 
2010).

In order to make sure that this shortcoming is not critical for our study we have repeated the inversions of
map A, using the HELIX++ inversion code (Lagg et al. 2004, 2009), which allows to consider both spectral lines 
simultaneously. The resulting histograms for velocity $V_{\rm LOS}$, magnetic field strength 
$\bar{B}$, inclination $\gamma$, etcetera, are very similar to those obtained from the 1-line inversions using
VFISV (see Figure 3). This makes us confident that our results are not affected by the fact 
that only one spectral line is being employed, or at least not to the point of invalidating our
results. We will return to this point in Section 5.

\section{Results}

In Figures 3a, 4a and 5a we present histograms for the number of pixels whose unsigned polarization signals
($|V|$ or $\sqrt{Q^2+U^2}$) have peak values $N$ times larger than or equal to the noise level in each map. 
This is will be referred to as SNR: signal-to-noise ratio. The vertical lines in these figures indicate the 
different SNR-thresholds that we have considered in our analysis: 3, 4.5 and  6. For instance, 
the vertical dashed-dotted line in Fig.~3a corresponds to SNR $=3$. This means that all pixels in map A, where 
\emph{at least one} of the polarization profiles (Stokes $Q$, $U$ or $V$) possesses a SNR $\ge 3$, are included in 
the histograms of the remaining panels and represented also by the dashed-dotted line. Likewise for SNR $\ge 4.5$ 
(solid lines) and SNR $\ge 6$ (dashed lines). The rest of the panels in Figs.~3-5 show the normalized histograms for the 
magnetic field strength $\bar{B}$ ({\bf b}-panels), magnetic field inclination with respect to the observer $\gamma$ 
({\bf c}-panels), and the magnetic filling factor $\alpha_{\rm mag}=1-\alpha_{\rm nm}$ ({\bf d}-panels), which is 
interpreted as the fractional area of the pixel filled with magnetized plasma. The normalization is done to ensure
 that the total area under each histogram equals to 1. In all these histograms we have avoided network regions by 
leaving out of the analysis those pixels where the mean field strength in the resolution element is larger than 750 Gauss. 

Note that in all {\bf a}-panels in Figs.~3-5 the histograms for the circular and linear polarization peak
at SNR $\approx 3$. This is due to photon noise. The probability that photon
noise produces a signal, at a given wavelength position, that it is above 3$\sigma_{\rm s}$ is only 0.3 \%. 
However, we must take into account that we are taking the peak value for all 112 (in circular polarization
or Stokes $V$) or 224 (in linear polarization or Stokes $Q$ and $U$; see Sect.~3.2), and therefore it is almost 
guaranteed that at every observed spatial pixel, photon noise will yield a polarization signal that is larger
or equal than 3$\sigma_{\rm s}$.

Since all considered maps are close to disk center, we can interpret the retrieved inclination with respect to the observer,
 $\gamma$, as the inclination with respect to the normal direction to the solar surface. A value of 
$\gamma=0\deg, 180\deg$ indicates a vertical (pointing upwards or downwards, respectively) magnetic field, whereas 
a value of $\gamma=90\deg$ indicates a magnetic field that is contained in the solar surface.

The resulting histograms are rather similar in all three (A,B and C) maps. For all SNR-thresholds considered,
the distributions of the average magnetic field strength $\bar{B}$, magnetic field inclination $\gamma$, and magnetic 
filling factor $\alpha_{\rm mag}$, peak at around 50-100 Gauss, 80-100 degrees and 20-30 \%, respectively. These results
involve the existence of rather weak and horizontal fields in the internetwork, occupying about 20\% of the resolution element 
at every pixel. This is in excellent agreement with the results presented in OS07a and OS07b. In particular, our retrieved 
histograms for Map A (Fig.~3) using a SNR-threshold of 4.5 are very similar to those in Figs.~3 and 4 in OS07a 
and Fig.~7 in OS07b.

\begin{figure}
\begin{center}
\includegraphics[width=9cm]{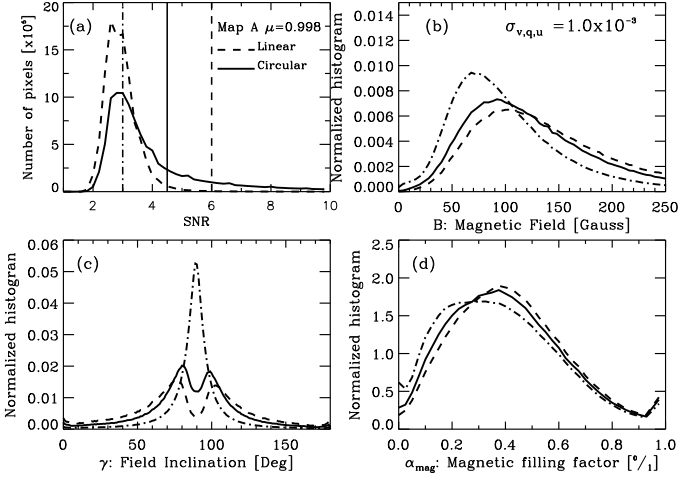}
\end{center}
\caption{Panel-{\bf a}: histogram for the number of pixels as a function of the signal-to-noise ratio
in the polarization profiles. Circular polarization (Stokes $V$) is indicated by the
 solid line, while the linear polarization ($\sqrt{Q^2+U^2}$) is shown in dashed. Panel-{\bf b}: histogram
for the magnetic field strength $\bar{B}$. Panel-{\bf c}: histogram for the inclination of the magnetic field 
vector $\gamma$. Panel-{\bf d}: histogram for the magnetic filling factor $\alpha_{\rm mag}$. In the last
three panels, the dashed lines display the histograms obtained from those pixels where any of three polarization
profiles has a SNR $\ge 3$, whereas solid and dashed lines represent SNR $\ge 4.5,6$, respectively. These 
thresholds are also indicated with vertical lines in panel a. This figure corresponds to map A, where the 
characteristic noise level is $\sigma_{\rm s} = 10^{-3}$.}
\end{figure}

\begin{figure}
\begin{center}
\includegraphics[width=9cm]{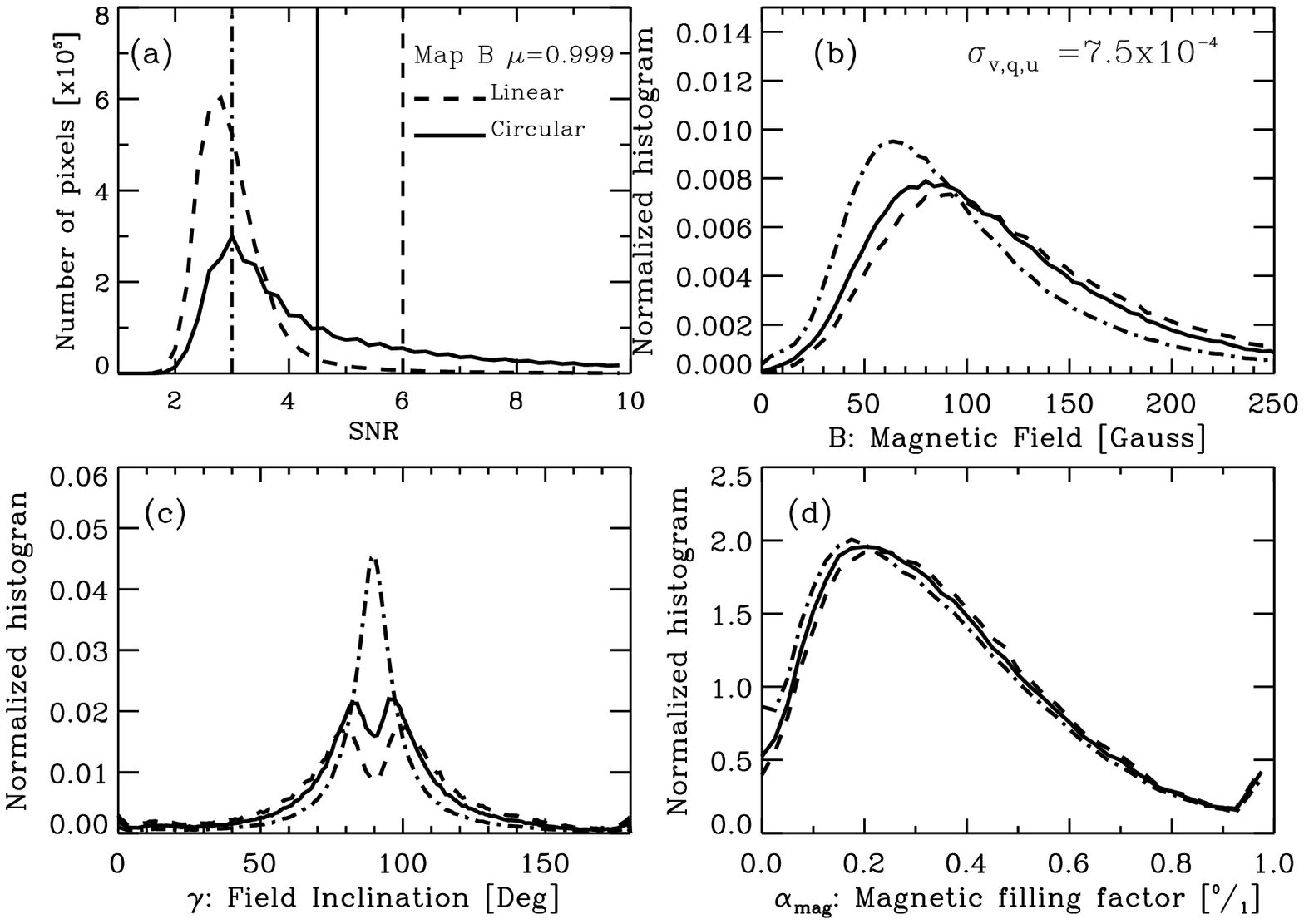}
\end{center}
\caption{Same as Fig.~3 but for map B, with a characteristic noise level of $\sigma_{\rm s} = 7.5\times 10^{-4}$.}
\end{figure}

\begin{figure}[ht]
\begin{center}
\includegraphics[width=9cm]{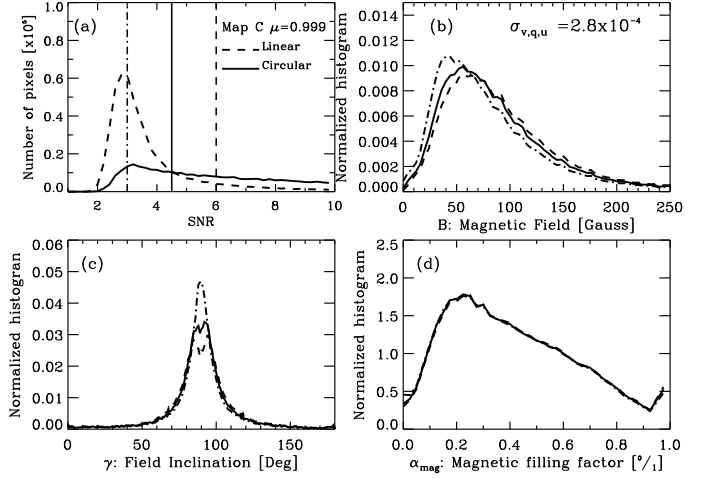}
\end{center}
\caption{Same as Fig.~3 but for map C, with a characteristic noise level of $\sigma_{\rm s} = 2.8\times 10^{-4}$.}
\end{figure}

A closer look to Fig.~3-5 reveals a couple of trends. The first one involves histograms for the same map 
(noise-level $\sigma_{\rm s}$ fixed) obtained with different signal-to-noise thresholds. Here we observe that increasing
the SNR-threshold yields histograms for the magnetic field strength and magnetic
filling-factor that shift towards higher values. In addition, the central peak at $\gamma=90\deg$ in the
histogram for the inclination becomes less pronounced and almost disappears. Still, the magnetic field still remains 
rather horizontal: $\gamma \in [60,120]\deg$. One can interpret this trend in terms of the Zeeman effect in the weak-field
regime, through which the amplitude of the polarization signals is proportional to the strength of the magnetic field.
From this point of view, imposing larger SNR-thresholds in the polarization immediately selects stronger fields.
The histograms for $\gamma$ (inclination) and $\alpha_{\rm mag}$ (magnetic filling-factor) indicate that these stronger 
magnetic fields tend to be more vertical and to occupy a larger fraction of the resolution element.

The second observed trend in the retrieved histograms is related to the map-to-map differences or, equivalently, 
to the noise level $\sigma_{rm s}$ in the four components of the Stokes vector. Comparing figures 3,4 and 5 reveals that the histogram
of the magnetic field strength shifts towards lower values from Map A ($\sigma_{\rm s}=10^{-3}$) to Map C ($\sigma_{\rm s}=
2.8\times 10^{-4}$). Again, this behavior can be explained as a consequence of being able to detect 
smaller polarization signals, hence weaker magnetic fields, when the noise is reduced. In addition, the 
disappearance of the peak at $\gamma=90\deg$ for large SNR-thresholds is much less pronounced in map C than
in maps A or B. An explanation for this will be offered in Section 6.

Figure 6 shows histograms for the longitudinal component of the magnetic field $B_\parallel=\bar{B}\cos\gamma$, and 
the transverse component of the magnetic field $\bar{B}_\perp=\bar{B}\sin\gamma$ for the three analyzed maps:
A (top panels), B (middle panels) and C (bottom panels). $B_\parallel$ and $B_\perp$ can be identified with 
the projection of the magnetic field vector on the vertical direction to the solar
surface and the projection of the magnetic field over the solar surface, respectively. Because the retrieved magnetic fields
are rather inclined ($\gamma\rightarrow 90\deg$), the horizontal component of the magnetic field $B_\perp$
is larger than the vertical component $B_\parallel$. 

We have calculated the center-of-gravity (CoG) of the histograms for $B_\parallel$ and $B_\perp$ in Fig.~6
and plotted them as a function of the signal-to-noise ratio in Figure 7. If we take 
as a reference the results obtained from considering only those pixels whose polarization levels 
are 4.5 times above the noise level: SNR$=4.5$, we then see that the CoG of 
$B_\perp$ (dashed lines) is located at around 122.0 Gauss for the Map A ($\sigma_{\rm s}=10^{-3}$; crosses), 
and it decreases to 87.4 Gauss when the noise level is $2.8\times 10^{-4}$ (Map C; diamonds). Differences
are even larger if we consider the peak-value (instead of the CoG) in the histograms: 90.6 Gauss in map A,
and 57.1 Gauss in map C. Meanwhile, the CoG values in the histogram of $B_\parallel$ (solid lines) 
only decreases from 21.6 (Map A; crosses) to 11.9 Gauss (Map C; diamonds). Similar trends are observed 
for other signal-to-noise thresholds. This indicates that both $B_\parallel$ and $B_\perp$ are responsible 
for the decrease in the CoG and peak value of $\bar{B}$ when the noise level is reduced (Figs~.3b-5b), with
$B_{\perp}$ being the main contributor.

Figs.~6 and 7 indicate that the CoG and peak values in the histograms for $B_\parallel$ and $B_\perp$
become smaller as the noise level decreases. In the light of this result, is it possible that, the current 
noise levels in the polarization signals (in all three maps considered) is responsible for the large values 
of $B_\perp$, and therefore for the peak at $\gamma=90\deg$ in the histograms for the inclinations ? 
And if so: is it conceivable that, given a low enough level of noise in the polarization signals, the detection 
thresholds for $B_\parallel$ and $B_\perp$ would be similar, thereby opening the possibility for the
detection of non-horizontal magnetic fields ? In the following we will address these two questions.

\section{Effect of the photon noise}

In order to study the effect that the photon noise has on the retrieval
of the inclination of the magnetic field vector in the quiet Sun, we have performed 
the following Monte-Carlo-like test: we first produce a large number of synthetic profiles (1024$\times$2047) 
where we assume that the magnetic field is always vertical. Half of the synthetic profiles posses $\gamma^{syn}=0\deg$ 
, while the other half have $\gamma^{syn}=180\deg$. The rest of the Milne-Eddington parameters needed for the synthesis: 
$\Delta_{\rm Dop}$, $V_{\rm LOS}$, $\eta_0$, $\alpha_{\rm nm}$ etcetera (see Sect.~3) are taken from the results
of the inversion of map A. This is done in order to employ kinematic and thermodynamic parameters that
are representative of the quiet Sun. The only exception is the magnetic field strength $\bar{B}$: for the synthesis 
we assume $\bar{B}$ to be equal to the vertical component of the field as obtained from the inversion of map A: 
$\bar{B}^{syn}=\bar{B}^{inv,A}\cos\gamma^{inv,A}$. Note that since $\gamma^{syn}=0,180\deg$ this ensures that
the vertical component of the magnetic field in this test is equal to the vertical component of
the magnetic field from map A: $B_{\parallel}^{syn}=B_{\parallel}^{inv,a}$.

We then proceed to add noise at different levels\footnote{We employ the random number generator algorithm
for normally distributed variables by Leva (1992).} and we employ the VFISV code (see Sect.~3) to invert the Stokes profiles
 of \ion{Fe}{I} 6302.5 {\AA} allowing all atmospheric parameters, including the inclination $\gamma$, to vary freely.
Finally, we compare the original distributions for $\gamma^{syn}$,
 $B_{\parallel}^{syn}$ and $B_{\perp}^{syn}$, with the retrieved ones. In our tests we have employed four different 
levels of noise: $\sigma_{\rm s} = 10^{-3}$ (as in map A), $7.5\times 10^{-4}$ (as in map B), $2.8\times 10^{-4}$ 
(as in map C) and finally, as an example of extremely high signal-to-noise observations, $\sigma_{\rm s} = 10^{-5}$. 
All numbers are expressed in units of the continuum intensity in the quiet Sun. As in Section 4, our analysis considers 
only those profiles where the signal-to-noise ratio in \emph{at least one} of the polarization profiles 
(Stokes $Q$, $U$ or $V$) is equal or larger than 3, 4.5 and 6.

\begin{figure}
\begin{center}
\includegraphics[width=9cm]{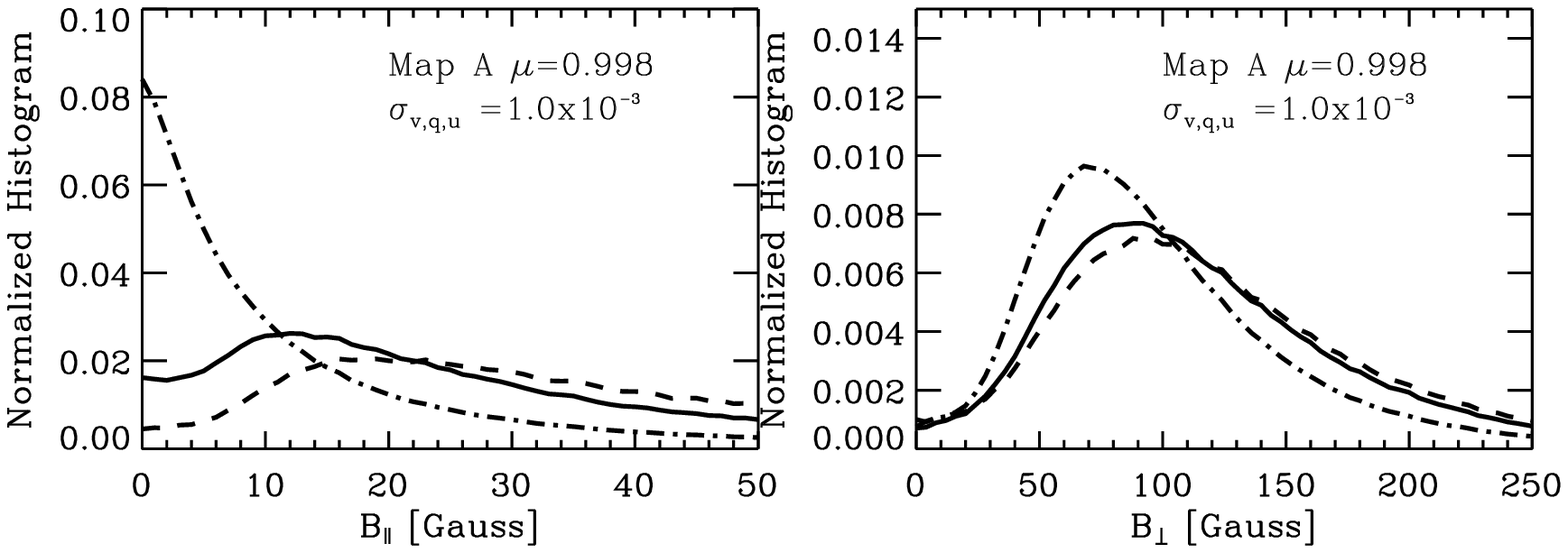} \\
\includegraphics[width=9cm]{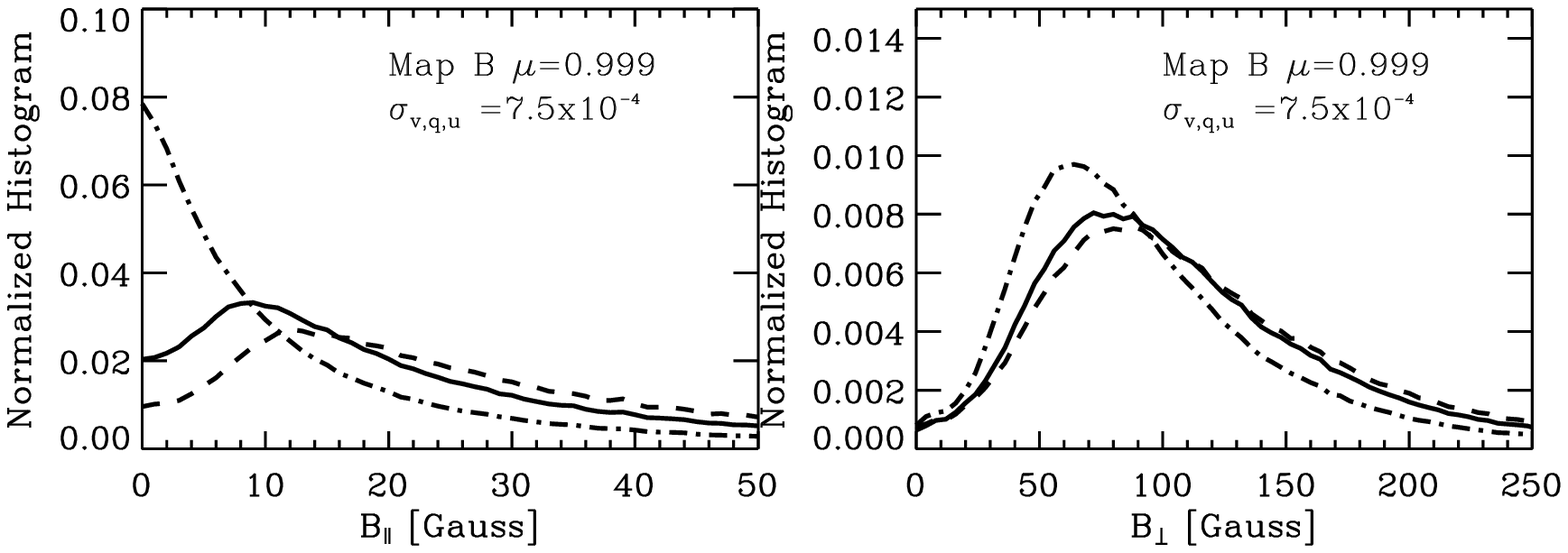} \\ 
\includegraphics[width=9cm]{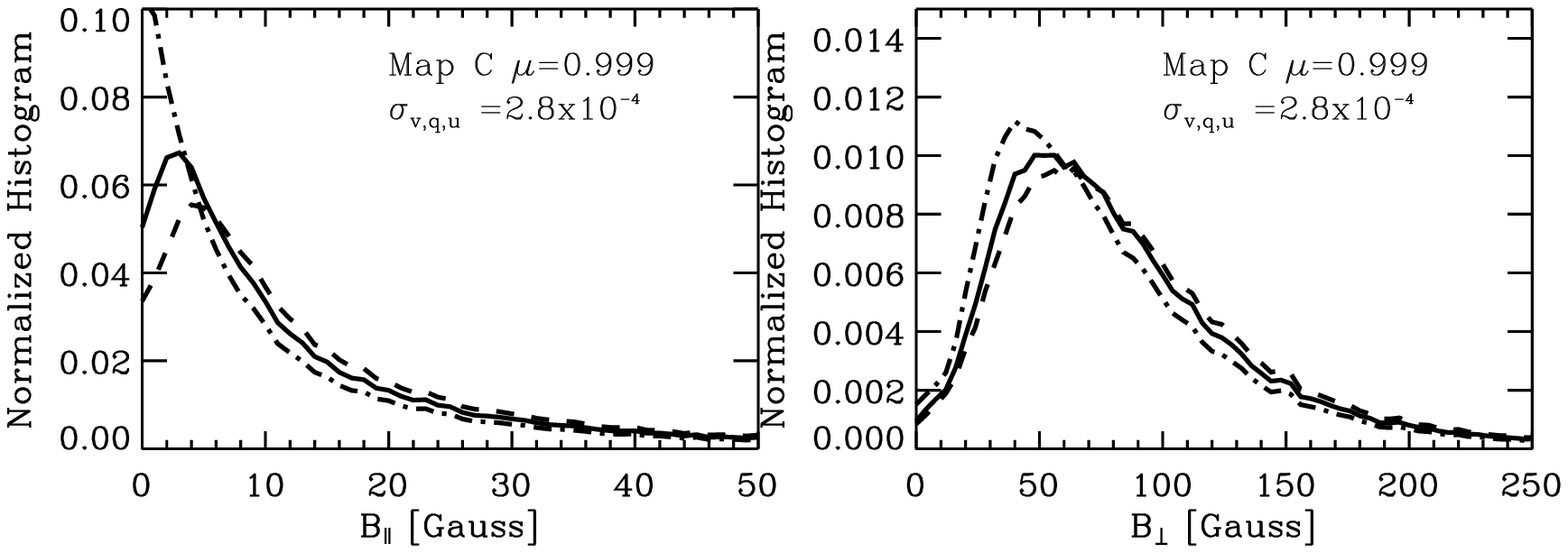}
\end{center}
\caption{Histograms for the longitudinal component of the magnetic field $B_\parallel$ (left panels) and
the transverse component of the magnet field $B_\perp$ (right panels). The first row corresponds to map A,
while the second corresponds to map B, and the third represents map C. Dashed-dotted, solid and dashed lines correspond 
to a SNR-threshold of 3, 4.5 and 6, respectively.}
\end{figure}

The results for the vertical component of the magnetic field $B_\parallel$ are presented in Figure 8. Each of the four panels in
this figure displays the results for the one of the four levels of noise under consideration. The original distribution
of vertical magnetic fields $B_{\parallel}^{syn}$ is indicated by the black lines. The retrieved distributions, corresponding
to the analysis of pixels where the SNR in at least one of the polarization signal is larger or equal than 3, 4.5 and 6,
are represented by the red, green and blue lines respectively. Note that a more realistic estimation of the original
 $B_{\parallel}$ is obtained by taken a lower SNR-threshold since, as already mentioned in Sect.~4, large SNR-thresholds tend to retain only the stronger
magnetic fields. This explains why the blue curves fail to obtain the correct distribution of vertical magnetic fields.
Only for the extreme case of $\sigma_s = 10^{-5}$ choosing different SNR-thresholds yields the correct
distribution of $B_{\parallel}$. It is important to notice that the behavior of the curves in Fig.~8 closely resembles
those in Fig.~6 for different SNR-thresholds obtained form the inversion of Hinode/SP data at different noise levels.

Figure 9 displays a similar plot to Fig.~8 but for the horizontal component of the magnetic field $B_\perp$. The original 
distribution, now a $\delta$-Dirac at $B_\perp=0$, is indicated by the vertical black arrow.
Here it can be seen that, despite having an original distribution of magnetic fields which was purely vertical, the retrieved horizontal
component of the magnetic field $B_{\perp}$ is much larger than the vertical component of the magnetic 
field $B_{\parallel}$ (see Fig.~8). The only exception to this is the case where $\sigma_s = 10^{-5}$, in which case the 
values of $B_{\perp}$ and $B_{\parallel}$ are of the same order.

An important feature of Figs.~8 and 9, which shows a strong resemblance with what we have seen in the real data 
(see Fig.~7), is that the CoG of the histograms (indicated by the vertical-dashed color lines) shifts 
towards lower values when the noise decreases. This happens for both $B_{\parallel}$  and $B_{\perp}$.

As it can be expected from these results, the retrieved inclination is mostly horizontal. This is demonstrated in Figure 10, 
where the histograms for the inclination of the magnetic field peak at around $\gamma \approx 90\deg$. This
happens in spite of having started with vertical fields, whose distribution is represented
in Fig.~10 by the two vertical black arrows at $\gamma=0,180\deg$. The fact that we have obtained horizontal fields
out of purely vertical ones is due to the sole effect of the photon noise. This is also true for noise levels as low as
$2.8\times 10^{-4}$, which corresponds to the the highest signal-to-noise data analyzed in OS07a, OS07b and LIT08.
Even at noise levels as low as $10^{-5}$ (lower-right panel in Fig.~10) the correct distribution of inclinations
is not properly recovered, although at least in this case, some amount of vertical fields remain after the inversion.

\begin{figure}
\begin{center}
\includegraphics[width=8cm]{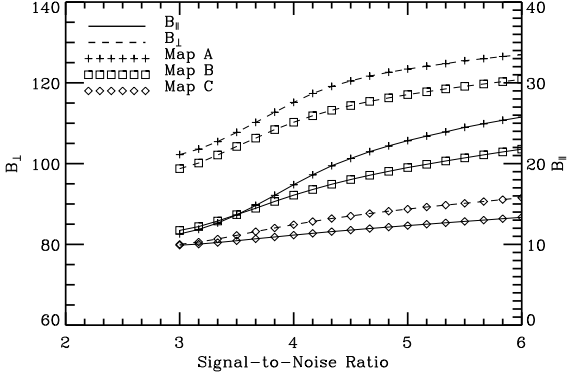}
\end{center}
\caption{CoG (center of gravity) of the histograms for $B_\parallel$ (solid) and $B_\perp$ (dashed) retrieved from 
the inversion of the profiles in map A (crosses; $\sigma_s=10^{-3}$), B (squares; $\sigma_s=7.5\times 10^{-4}$) and C 
(diamonds; $\sigma_s=2.8\times 10^{-4}$), as a function of the SNR-threshold.}
\end{figure}

Another interesting feature of our Monte-Carlo simulation is that it also reproduces the drop in the central peak at $\gamma=90\deg$
when using a SNR-threshold of 4.5 and 6 or, in order words, when only considering pixels where the polarization signal 
are 6 times or more above the noise level:. This was already seen from the inversions of real data:  Figs.~3c-5c. However,
this simulation fails to reproduce the fact that the observed drop at $\gamma \approx 90\deg$ is less pronounced for smaller
noise levels.

The tests presented in this section demonstrate that it is possible to deduce the existence of horizontal fields from
purely vertical ones due to the sole effect of photon noise. Photon noise is also responsible for an overestimation
in $B_\perp$ and therefore also in $\bar{B}$. These effects occur for noise levels as low as $\sigma_{\rm s}=10^{-5}$.
Therefore, results from this simulations cast doubt as to whether it is possible to correctly retrieve the distribution of the
magnetic field vector in the quiet Sun internetwork. This will be addressed in the next section.

Finally it is important to mention a couple of consistency checks that have been carried out. The first one was
performed in order to study if our results are biased as a consequence of the inability of the VFISV inversion code to 
invert both \ion{Fe}{I} 6301.5 {\AA} and \ion{Fe}{I} 6302.5 {\AA} spectral lines (see Sect.~3.2), we have repeated our
 Monte-Carlo experiments using the SIR inversion code (Ruiz Cobo \& del Toro Iniesta 1992) and inverting both spectral 
lines simultaneously. Due to SIR's limited speed we have run the experiment with only 5000 profiles, but this is sufficient 
to confirm the results presented throughout this section. The second consistency check is to repeat these experiments
in the absence of photon noise. With this test we have confirmed that vertical magnetic fields are actually retrieved 
(initial $\delta$-Dirac distributions), and to test the convergence algorithm of the inversion code.

\section{Selection effects}

In the inversion of observed Hinode/SP data (Sect.~4) we have applied a selection criteria that considers in the analysis 
all those pixels where \emph{any} of the polarization signals ($Q$, $U$ or $V$) are above a certain signal-to-noise ratio. This 
is the same criteria applied in the analysis of simulated data in Sect.~5 and also in OS07a and OS07b.

A very important consequence to be drawn from Sect.~5 is that the distributions for the magnetic field vector 
obtained from the inversion of the simulated data, are not trustworthy as long as Stokes $Q$ and $U$ are dominated by 
noise. This was clearly the case in the inversion of the simulated data in that section since there it had been assumed 
that the magnetic field was vertical.

It is therefore critical to study whether the aforementioned criteria, when applied to Hinode/SP data, select
 pixels where the linear polarization profiles ($Q$ and $U$) are above the noise or, on the contrary, where 
$V$ is the only Stokes parameter above the noise. For the simulated data in Sect.~5, the latter was certainly 
the case since the signals in $Q$ and $U$ were due only to photon noise. What about the observations from Hinode/SP 
employed in Sect.~4 ?

In Figure 11 we show the percentage of pixels (with respect to the total number of pixels in each Hinode/SP map)
as a function of the SNR-threshold applied. The three maps with different noise levels considered in Sect.~4
are indicated by the different symbols: crosses (map A: $\sigma_{\rm s}=10^{-3}$), squares (map B: $\sigma_{\rm s}=7.5\times 10^{-4}$),
and diamonds (map C: $\sigma_{\rm s}=2.8\times 10^{-4}$). Here we have applied two different criteria. The first one is 
the same as in Sect.~4 and 5: a pixel is taken into account if \emph{any} of three polarization profiles ($Q$, $U$ or $V$) 
has a given signal-to-noise ratio (solid lines). The second one considers only those pixels were either $Q$ or $U$ posses
a certain signal-to-noise ratio (dashed lines).

Taking as a reference a SNR-threshold of 4.5 (same as in OS07a) we find that 30.2 \% of the pixels contained in map A
posses at least \emph{one} of polarization signal that is above this threshold. However, if we restrict our analysis to $Q$ or $U$ 
only, we realize that only 3.4 \% of the pixels fulfill this criteria. This implies that the vast majority of the pixels
analyzed have linear polarization profiles that are dominated by noise and thus, as demonstrated in Sect.~5, this automatically
results into horizontal fields (see Fig.~10). Consequently, the peak at $\bar{B}=100$ G and $\gamma=90\deg$ in Fig.~3 in this work 
and also in Fig.~3 in OS07a is not real but due to noise.

In map C, with a much lower noise level than map A, up to 80.0 \% of the pixels posses signals where at least one of the polarization
profiles is above 4.5 times the noise level. However, the number drops to 31.6 \% if we consider only $Q$ or $U$. As a
consequence, only $\approx 1/3$ of the peaks at $\bar{B}=50$ G and $\gamma=90\deg$ in Fig.~5 is real, with the remaining $\approx 2/3$ being 
due to noise. This explains why in map A the peak at $\gamma \approx 90\deg$ disappears for large SNR-thresholds: it is mostly noise 
(Fig.~3c), while remaining in map C (Fig.~5c): part of the peak there is actually real.

What is then the inclination and strength of the magnetic field vector in the pixels where $Q$ and $U$ are not above the noise level ?
This represents 97 \% of all pixels included in the analysis of Fig.~3 (map A) and about 70 \% of all pixels included in the analysis
of Fig.~5 (map C). As demonstrated in Sect.~5 it is possible that they are vertical $\gamma \approx 0\deg$ with a field strength of $\bar{B}
\approx 10$ G. However, as long as $B_\perp$ remains below our detection threshold this is only speculation.

In order to improve the reliability of the histograms for the magnetic field vector it is possible to repeat the analysis
in Sect.~4 but employing only those pixels where the linear polarization signals (Stokes $Q$ or $U$) are 4.5 times above the noise level.
In this case we still obtain the central peak at $\gamma=90\deg$ and $B=60-120$ G in all three maps considered. This was to be expected
because, $Q$ and $U$ signals sufficiently above the noise implies the existence of
highly inclined magnetic fields ($\gamma\rightarrow 90\deg$). This is so because in the weak-field regime, the 
linear polarization is a second order effect (Landi Degl'Innocenti, 1992) in the magnetic field as compared to the
circular polarization.

\begin{figure}
\begin{center}
\includegraphics[width=9cm]{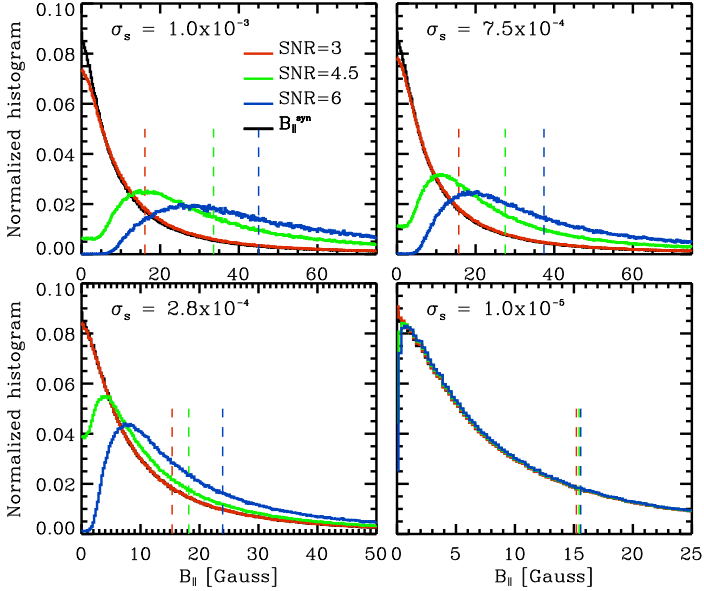}
\end{center}
\caption{Histograms for the original longitudinal component of the magnetic field $B_\parallel^{\rm syn}$ used in the
synthesis of the simulated data (black). The retrieved distributions from the inversion are indicated by the color
lines: SNR $\ge 3$ (red), SNR $\ge 4.5$ (green) and SNR $\ge 6$ (blue). The CoG of the retrieved histograms
are indicated by the vertical dashed lines. Each panel shows the results for different levels of noise: $\sigma_s=10^{-3}$
(as in observed map A; upper-left), $\sigma_s=7.5\times 10^{-4}$ (as in observed map B; upper-right), $\sigma_s=2.8\times 10^{-4}$
(as in observed map C; lower-left), and finally $\sigma_s=10^{-5}$.}
\end{figure}

\begin{figure}
\begin{center}
\includegraphics[width=9cm]{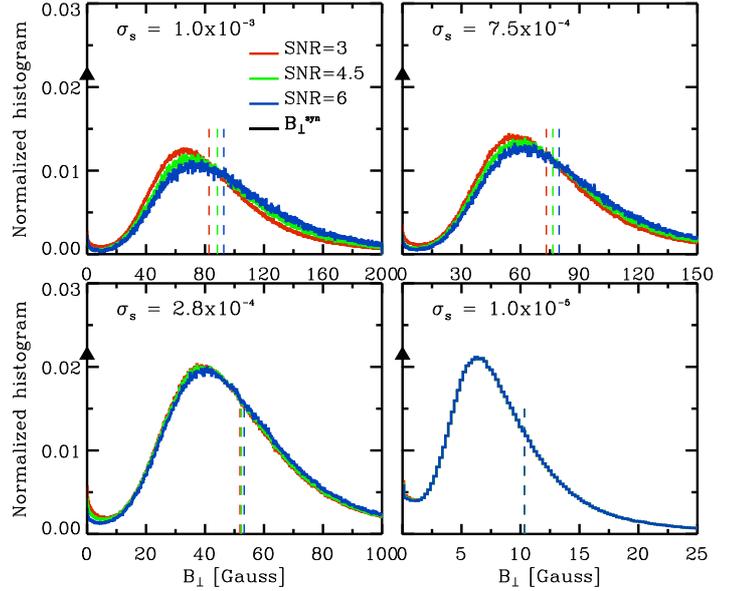}
\end{center}
\caption{Same as Fig.~8 but for the transverse component of the magnetic field $B_\perp$. The vertical black arrow
indicates the original distribution employed to produce the synthetic Stokes profiles: it's a $\delta$-Dirac centered
at 0 because the synthetic profiles were obtained assuming that the magnetic field is longitudinal/vertical ($\gamma=0$)
and therefore $B_\perp^{\rm syn}=0$. Note however, that the effect of the noise induces the inversion code to retrieve 
large values for the transverse component of the magnetic field (blue and red lines).}
\end{figure}

\begin{figure}
\begin{center}
\includegraphics[width=9cm]{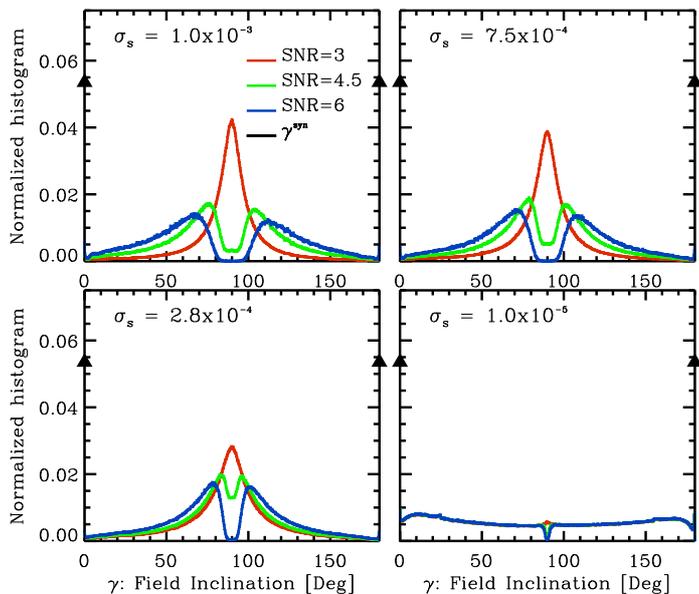}
\end{center}
\caption{Same as Fig.~8 but for the inclination of the magnetic field vector $\gamma$. The vertical black arrows
indicates the original distribution employed to produce the synthetic Stokes profiles: it's a double $\delta$-Dirac centered
at 0$\deg$ and 180$\deg$ because the synthetic profiles were obtained assuming that the magnetic field is longitudinal/vertical. 
Note however, that the effect of the noise induces the inversion code to retrieve highly inclined magnetic fields (color lines).}
\end{figure}

However, it is important to clarify that histograms obtained applying the selection criteria to
the linear polarization would only contain a minority of internetwork pixels, leaving out many of them (up to 90 \% 
of internetwork pixels in map A or up to 70 \% of pixels in map C) whose inclination is unknown. Our tests show that
in order to correctly infer the inclination of those pixels where only Stokes $V$ is above the signal, a noise level
better than $\times 10^{-5}$ (see Fig.~10) is needed.

According to Fig.~11 one might think that, even though only 3.4 \% of the pixels in map A and 31.6 \% in map C, posses
a signal-to-noise larger or equal to 4.5, these numbers grow up to 69.6 \% and 85.9 \% respectively if a SNR-threshold
of 3 is considered. However, as explained in Sect.~4, SNR $\ge 3$ is not sufficient to reliably infer the magnetic field
vector because photon noise produces signals that are above 3 times the noise level in most of the pixels.

\section{Conclusions}

In this paper we have studied the distribution of the magnetic field vector (strength $\bar{B}$ and inclination with respect
to the vertical direction $\gamma$) in internetwork regions with data from the spectropolarimeter (SP) on-board the 
Japanese satellite Hinode. We have selected three different maps recorded at disk center with different integration
 times, and therefore different noise levels. In all three cases, the retrieved inclination of the magnetic field peaks 
at $\gamma=90\deg$, in agreement with previous results (Orozco Su\'arez et al. 2007a, 2007b).

In addition, we have found that the distribution of vertical/longitudinal fields ($B_\parallel=\bar{B}\cos\gamma$)
and horizontal/transverse fields ($B_\perp=\bar{B}\sin\gamma$) shift towards smaller values when we analyze data
with lower levels of photon noise. In spite of this decrease, the relation $B_\perp >> B_\parallel$ holds, which explains
why all analyzed maps yield histograms for the inclination with a clear peak at $\gamma=90\deg$.

\begin{figure}
\begin{center}
\includegraphics[width=9cm]{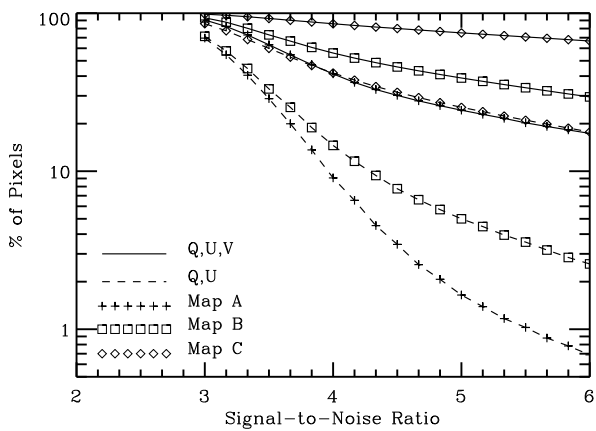}
\end{center}
\caption{Number of pixels (with respect to the total amount of pixels in each map) as a function of their signal-to-noise ratio.
In solid lines it is showed the number of pixels when considering any of the polarization profiles (Stokes $Q$, $U$ or $V$),
whereas in dashed lines it is plotted the number of pixels when considering only the linear polarization (Stokes $Q$ or $U$).
Symbols indicate each of the three maps employed: crosses (map A; $\sigma_s=10^{-3}$), squares (map B; $\sigma_s=7.5\times 10^{-4}$ 
, and diamonds (map C; $\sigma_s=2.8\times 10^{-4}$.}
\end{figure}

To understand this result we have carried out a Monte-Carlo-like simulation in which we produce a large number
of synthetic profiles assuming vertical fields ($\gamma=0,180\deg$) and, after adding photon noise at different levels, 
we invert the synthetic Stokes profiles to investigate whether the original distributions ($\delta$-Dirac functions) 
can be recovered. These tests show that, peaks at $\gamma=90\deg$ (horizontal fields) are obtained for noise levels 
as low as $\sigma_s=2.8\times 10^{-4}$ even though in the synthesis we had used only vertical ones. This means that
 the histograms for the inclination obtained with real data are contaminated to a large extent by the effect of 
the photon noise, which makes vertical and horizontal fields indistinguishable.

To what extent the histograms are contaminated depends on the selection criteria employed to select pixels
that enter in the analysis. If we select pixels where the signal-to-noise ratio in any of the polarization profiles 
(Stokes $Q$, $U$ or $V$)  is above 4.5, the contamination can be as large as 70-95\% (depending on the noise
level). This renders the inferred distributions for the magnetic field inclination and transverse component of the magnetic
field, meaningless. This is the approach followed in Orozco Su\'arez et al. (2007a, 2007b). Other authors such as 
Lites et al. (2007) and Kobel (2009) impose a larger threshold for Stokes $V$ (SNR $\ge 6$) than for Stokes $Q$ and $U$ 
(SNR $\ge 2$), which can partially avoid the contamination.

A better approach is to analyze only those pixels where the linear polarization profiles (Stokes $Q$ and $U$) have
a signal-to-noise ratio larger than or equal to 4.5. This strategy has the advantage that the inclination of the
magnetic field can be properly retrieved by the inversion of the radiative transfer equation applied to the observed 
Stokes profiles. A similar conclusion is reached by Asensio Ramos (2009). The main shortcoming of this selection
 procedure is that it considers only small regions within the internetwork: 5-30 \% (depending on the noise level), 
leaving out large portions where the inclinations is not determined. We conclude therefore that the real distribution 
of inclinations in the quiet Sun internetwork remains unknown. A proper inference can only be carried out with noise 
level that are beyond the current capabilities of modern spectropolarimeters. Workarounds can be found by introducing 
temporal averages, but the required integration time exceeds the lifetime of granulation. It is therefore not clear 
what is being represented by histograms obtained with averages spanning several minutes.

The importance of the photon noise has been already acknowledged by Asensio Ramos (2009) and Stenflo (2010).
In the latter work, results are interpreted in terms of an isotropic distribution
of magnetic fields. However, these works do not analyze low-noise data, thus missing the
contribution from pixels with strong $Q$ and $U$ signals: up to 30 \% of the total amount of pixels. 
The question is therefore whether an isotropic distribution of fields can explain these regions of 
truly horizontal fields.

Asensio Ramos (2009; see also Mart{\'\i}nez Gonz\'alez et al. 2008) also conclude that the distribution of 
magnetic fields is quasi-isotropic, however, whereas the latter refers to pixels where the polarization signal
is above the noise level, the former infers a quasi-isotropic distribution only for those pixels whose polarization signal
is below the noise. Our work differs from these ones in that in our case we can explain many features of 
the resulting histograms for the inclination employing only vertical fields. In the future we intent to
adress the possibility of isotropism in more detail.

\begin{acknowledgements}
We would like to thank Dr. Bruce W. Lites, Dr. Luis Bellot Rubio and Dr. David Orozco Su\'arez for fruitful 
discussions in the subject. We are also grateful to Dr. Andreas Lagg for comparing inversion results using
the HELIX++ code with the results from the VFSIV code. This work analyzes data from the Hinode spacecraft. 
Hinode is a Japanese mission developed and launched by 
ISAS/JAXA, collaborating with NAOJ as a domestic partner, NASA and STFC (UK) as international partners. 
Scientific operation of the Hinode mission is conducted by the Hinode science team organized at ISAS/JAXA. 
This team mainly consists of scientists from institutes in the partner countries. Support for the post-launch 
operation is provided by JAXA and NAOJ (Japan), STFC (U.K.), NASA, ESA, and NSC (Norway). This work has also
made use of the NASA ADS database.
\end{acknowledgements}

\end{document}